# Anonymous Communication in Peer-to-Peer Networks for Providing more Privacy and Security

Ehsan Saboori and Shahriar Mohammadi

*Abstract*—One of the most important issues in peer-to-peer networks is anonymity. The major anonymity for peer-to-peer users concerned with the users' identities and actions which can be revealed by any other members. There are many approaches proposed to provide anonymous peer-to-peer communications. An intruder can get information about the content of the data, the sender's and receiver's identities. Anonymous approaches are designed with the following three goals: to protect the identity of provider, to protect the identity of requester and to protect the contents of transferred data between them. This article presents a new peer-to-peer approach to achieve anonymity between a requester and a provider in peer-to-peer networks with trusted servers called suppernode so that the provider will not be able to identify the requester and no other peers can identify the two communicating parties with certainty. This article shows that the proposed algorithm improved reliability and has more security. This algorithm, based on onion routing and randomization, protects transferring data against traffic analysis attack. The ultimate goal of this anonymous communications algorithm is to allow a requester to communicate with a provider in such a manner that nobody can determine the requester's identity and the content of transferred data.

## I. INTRODUCTION

A peer-to-peer network is a dynamic and scalable set of computers (also referred as peers). The peers can join or leave the network at any time. [1] The basic idea of a peer-to-peer network is to build a virtual layer over the application or network layer. In such an overlay network all peers interconnect with each other. All peers are both the resource consumers and providers. Currently, file-sharing is the most popular application in peer-to-peer systems [2]. Peer-to-peer networks can be divided into structured and unstructured classes. Structured peer-to-peer networks map each peer as well as the index information of each resource into a globally position such as Distributed Hash Table (DHT) in a highly organized structure. This paradigm has two main drawbacks which limit the implementation in real world. First, it cannot support the fuzzy query and Second, the DHT structure has large overhead to individual peers and too difficult to maintenance.

In Unstructured peer-to-peer networks, peers can join and leave networks simply and there are not any structured patterns there. This article focuses on the unstructured peer-to-peer networks because this kind of network is better to implement and provide security and anonymity.

There are three different roles that each peer can play in peer-to-peer networks: a *provider* (also called a responder, host or publisher) to provide services upon requests, a *requester* (also called an initiator) to request services, and a *proxy* (also called an intermediate peer) in which routs data from a peer to another peer. According to these roles there are three aspects of anonymity in peer-to-peer networks: *Provider anonymity* that hides the identity of a provider against other peers, *Requester anonymity* that hides a requester's identity and *Mutual anonymity* that hides both provider's and requester's identities. In the most stringent version, achieving mutual anonymity requires that neither the requester, nor the provider can identify each other, and no other peers can identify the two communicating parties with certainty. [3]

The proposed algorithm in this article provides requester anonymity to protect the identity of the requester and the transferred data against other peers specially the intruders. The proposed algorithm is based on *Onion Routin* mechanise. Onion Routing is the technique in which the requester and the provider communicate with each other anonymously by means of some intermediate peers called as onion routers. In this technique, messages route between onion routers. The messages encrypted with onion router's public key. Each onion router learns only the identity of the next onion router.

## II. UNSTRUCTURED PEER-TO-PEER NETWORKS

Unstructured peer-to-peer networks classified into three classes: centralized, decentralized, and hybrid. A centralized unstructured peer-to-peer networks, holds one or more centralized servers to provide resource index services. Those servers maintain index lists of available resources of all peers. Each peer sends requests for desired resources, called queries, to the index servers. For each query, index servers search in the maintained index lists and reply a result to the requesting peer. The response includes the description of resources and providers' IP addresses. Upon responses, the requesting peer chooses a desired provider and directly contacts it to gets services. [2] Fig. 1 illustrates the centralized unstructured peer-to-peer networks. Centralized peer-to-peer benefits from the efficient search performed by index servers. However, the overt drawback is that index servers are vulnerable to single point of failures and denial of service attacks. [1]

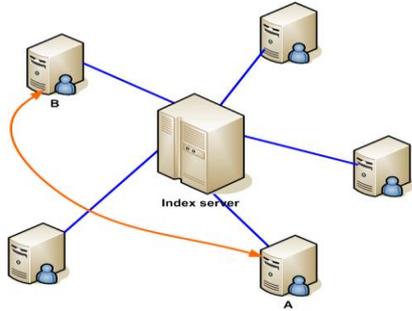

Fig. 1. A centralized unstructured peer-to-peer networks

Decentralized unstructured peer-to-peer networks remove index servers and Instead of processing queries in a centralized manner, peers usually employ a flooding mechanism to issue queries. Each requesting peer broadcasts a query to its neighbouring peers. The query is broadcast and rebroadcast in the system, which is called a flooding procedure. Each peer caches a local routing table for relaying queries. If a peer within the flooding scope has a matched object, it becomes a provider. All providers deliver their responses to the requesting peer. Each response is delivered along the reversed path of the query message until it reaches the requesting peer. The requester then selects a desired provider and directly gets services from the chosen provider. To keep the flooding scalable, a TTL (Time-To-Live counter) value is set in the query message to constrain the hops it traverses. The decentralized model is more reliable than the centralized model due to the elimination of centralized servers. However, the flooding search incurs a large amount of traffic cost and degrades the search efficiency. Fig. 2 shows a decentralized unstructured peer-to-peer networks.

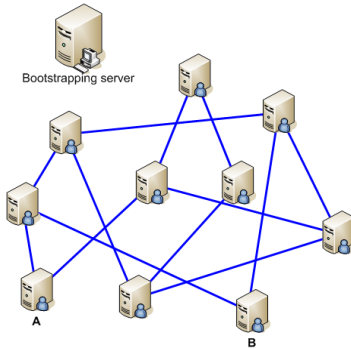

Fig . 2. A decentralized unstructured peer-to-peer networks

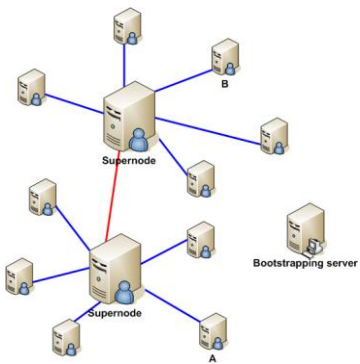

Fig. 3. A hybrid unstructured peer-to-peer networks

Combining the advantages of the centralized and decentralized models, the hybrid unstructured peer-to-peer model improves the search efficiency while maintaining the reliability. The hybrid unstructured peer-to-peer comprises a larger number of small groups. Each group is a small centralized peer-to-peer system, with a group leader, called *suppernode*, behaving as the index server for other group members. All *suppernode* are organized into a decentralized overlay. Fig. 3 shows a hybrid unstructured peer-to-peer networks.

### III. EXISTING ANONYMITY PROTOCOLS

**Crowds** [4] are an anonymous web transaction protocol and one of the oldest anonymizer networks and only provide requester anonymity. Crowds contains a closed group of participating nodes called *jondos* and uses a trusted third party as centralized crowd membership server called *blender.* The new *jondo* requests crowd membership from the *blender,* then the *blender* replies with a list of all current crowd members. After that, the *blender* informs all previous members of the new member. The requester node selects randomly a *jondo* from the member list and forwards the request to it. The following nodes decide randomly whether to forward the request to another node or to send it to the server. Crowds is vulnerable to DoS attacks.

**Hordes** [5] provides requester anonymity by adopting the Crowds probabilistic forwarding mechanism, and achieve provider anonymity by performing a multicast transmission. Since the replying path is the shortest multicast path from the provider to the requester, Hordes significantly reduces the response time. However, peers in Hordes must participate in the multicast relaying, which incurs a huge traffic and wastes the bandwidth.

**Freenet** [6] is a searchable peer-to-peer system for censorship resistant document storage. It is both an original design for anonymity and an implemented system. While it does not aim to hide the provider of a particular file it does aim to make it impossible for an attacker to find all copies of a particular file. A key feature of the Freenet system is that each node will store all the files that pass across it, deleting the least used if necessary. A hash of the title (and other key words) identifies the files. Each node maintains a list of the hashes corresponding to the files on immediately surrounding nodes. A search is carried out by first hashing the title of the file being searched for, and then forwarding the request to the neighboring node that has the file with the most similar hash value. The node receiving the request forwards it in the same way. If a file is found, it is sent back along the path of the request. This unusual search method implements a node-to-node broadcast search one step at a time. Over time it will group files with similar title hash values, making the search more efficient.

**Tor** [7] is an advanced version of Onion Routing. Instead of using a single layered encryption packet, say an onion, Tor implements an incremental path construction in which the initiator extends the path hop by hop and negotiates session keys with each intermediate node on the path. As a benefit, the anonymous transmission is more reliable since the intermediate nodes on the path are online after the path construction. Tor is more convenient than Onion Routing in supporting TCP-based applications.

**Tarzan** [8] provides a best-effort delivery service over IP layer. Each node of Tarzan is based on fundamental path-based technique to anonymously deliver messages. Different from onion routing which only provides a small proxy set, each Tarzan peer involves all other nodes in its proxy set. To accomplish this, Tarzan uses a gossip-based protocol for proxy discovery. The most elegant design in Tarzan is to inject mimic traffic to communication links to protect real data flows against eavesdropping. In Tarzan's topology, each node establishes k bidirectional links with k neighbours. All nodes maintain and balance the mimic traffics according to a number of criteria to shape the traffic into a time-invariant pattern. This defends the real traffic against being distinguished from the mimic ones. However, Tarzan's architecture is insufficient to guarantee a rapid flux in peer-to-peer systems. Tarzan's proxy discovery scheme and key exchange mechanism also incur significant amount of traffic.

## IV. ATTACKS AGAINST ANONYMOUS NETWORKS

The attackers may be system members or intruders from outside. The ultimate target is to locate the requester, provider, and find what they are transferring. [1] The topology of the peer-to-peer network is very important, because it provides very crucial and vital information about the system to attacker to compromise the networks. *Time-to-Live Attacks*, *Statistical Attacks, Denial of Service Attacks* and *traffic analysis* are kinds of popular attacks which attackers use to compromise the identity of provider and requester.

*Time-to-Live Attacks*: Time-to-live counters determine the maximum number of hops for a message and are used in most peer-to-peer networks to avoid flooding. If an attacker can send a request to a node with such a low time-to-live counter that the packet will probably not be forwarded, any response relieves that note as the provider. [9]

*Denial of Service Attacks*: A peer-to-peer networks cannot be used for anonymous if it cannot be used at all. Denial of service attacks can be particularly awkward when nodes can act anonymously, as this could mean that the node performing a Denial of service attack could not be identified and removed from the system. While anonymous systems cannot stop all Denial of service attacks. [9]

*Statistical Attacks*: Any attacker will be able to get statistical information over a period of long time. Networks may probably safe for a single run but may reveal information about the identities of their peers when all the observable messages of a longer run are analysed for patterns.

*Traffic analysis*: Making use of the traffic data of a communication to extract information. *Interception* and *cryptanalysis* are two techniques to analyse the transferred data. The reliable anonymous approach must invulnerable against these kinds of attacks.

## V. PROPOSED DUAL-PATH TECHNIQUES FOR REQUESTER ANONYMITY

We present our algorithm for achieving requester anonymity with the help of trusted third party called suppernode that only keeps network's map. Each peer must send a trigger signal to suppernode either periodically or when it wants to join/leave the network.

As we mentioned before, the ultimate goal of anonymous peer-to-peer networks is to hide the user identities, such as the user's ID and IP address. In fact, anonymity can be regarded as a special encryption on the messages to conceal correlations between the messages and the senders. The anonymizing process is performed during publishing, communicating, searching, and retrieving. Therefore, protecting the messages in communication is essential for anonymity.

The proposed algorithm is the way that the requester can connect to arbitrary provider and transfer data with it, so that any peers such as provider cannot detect the requester's identity. The basic principle is relay messages from requester to provider through multiple intermediate peers so that the true origin and destination of the messages is hidden from other peers. The requester creates a dual-path which contains a path to send request and another to get respond from provider so that the provider cannot compromise the requester's identity. The transferred data between requester and provider is encrypted to protect it against eavesdropping. So, in this algorithm there are two paths to connect requester to provider: *request path* and *response path*. Both of them are initiated by requester randomly. The requester can change these paths randomly while connecting to provider at any time. Fig. 4 illustrates a request and response paths in the network.

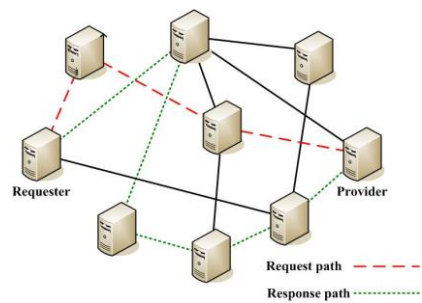

Fig. 4. Dual-Path paradim (The base principle of the request and response paths)

Each peer must join the network for getting services. The new peer requests a list of peers in the networks from the suppernode. The suppernode replies with a list of all current peers. After that suppernode informs all peers of the new member. For leaving the networks, the peer must send a removing signal for informing the suppernode that it wants to leave there. Then, the suppernode updates the list of peers furthermore it announces other peers automatically. Each peer must send a trigger signal to the suppernode periodically, to inform the suppernode that it is alive. After a period of time, if the suppernode does not sense any trigger signal from a peer, it will remove the peer from the list.

## VI. CREATE ANONYMOUS DUAL-PATH BY REQUESTER

The most important and vital part of the proposed algorithm is creating optimal anonymous dual-path for transferring data via them. As it is mentioned before, suppernode has a list of peers in the network. A requester sends a signal to the suppernode and requests a list of all current peers. The suppernode replies to the peer and sends a list of the live peers in the networks. Now the peer has a map of the network and can create dual-path base on this information. The requester chooses two sets of peers randomly. One of them is used for request and the other is used for response. The requester, requests via the request path and provider, responds via the response path. The response path is embedded in request path by the requester and when the provider gets a message from the requester it knows which peer must be given the response message as the first. For this purpose, the requester creates a wrapped message (Has been shown in section) and sends it to the next peer. The next peer decrypts the message and sends it to the next peer which is determined in the message. This process is continuing until the message is being received by the provider. When the provider wants to response the request, it sends the respond message to the peer that determines in the tail of the received message from the requester. The next peer does the same action until the massage is being received by the requester. Let's consider peers $P_1$, $P_2$ and $P_3$ which are chosen randomly by requester for request path and $P_4$, $P_5$ and $P_6$ which are chosen for response path. Also consider M, the message, which the requester wants to send. Fig. 5 shows the Dual-path created by requester. In this figure, "A" acts as a requester and "B" acts as a provider. "A" creates two paths to communicate with "B" and sends messages via them. "A" must rely messages through $P_1$, $P_2$ and $P_3$ (request path) to send them to provider. Also "A" receives the response of its request through $P_4$, $P_5$ and $P_6$ (response path).

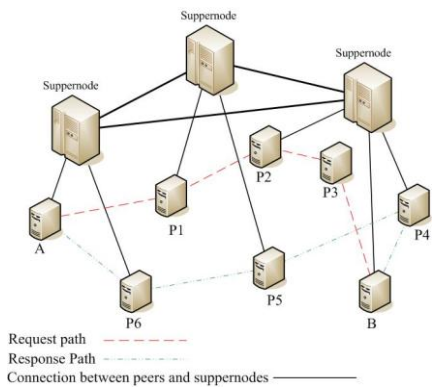

Fig. 5. Dual-Path between A and B as requester and provider

After the requester (A) creates the Dual-Path, now it must create the packet of the messages. To create the packets, the requester (A) must encrypt the messages by intermediate peers' public keys in a layer by layer structure, such as onion routing [4] mechanism. Fig. 6 shows how the requester wraps the message by intermediate peers public keys.

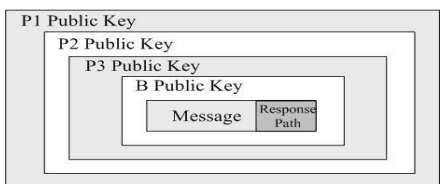

Fig. 6. Use intermediate peers' public keys to wrap message

While the requester wraps the message, it embeds the response path in the end of message as it is shown in Fig. 6. This part of message contains the response path. The structure of response path in wrapped message is illustrated in Fig. 7.

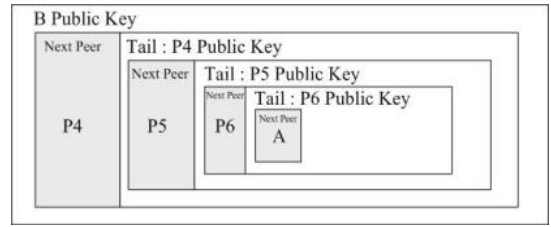

Fig. 7. The structure of response path in wrapped message

When the provider (B) receives the packets, it extracts the message and the response path packet. Each response packet has two parts, the "*Next Peer*" and the "*Tail*". The "*Next Peer*" part contains the next peer in which the message must be sent to it. After extracting the response path packet by provider (B), it encrypts the response message by P4 public key and attaches the "Tail" part of response path packet at end of it. Now the provider (B) sends the wrapped message to P4. P4 does same process and sends the received messages to P5. P5 sends the messages to P6 and at last P6 sends the messages to the requester (A). When the requester received its response, one dual-path cycle is completed and the requester can use this dual-path or choose another dual-path for more security in order to transfer the messages.

## VII. USE PRIVATE KEY FOR MORE EFFICIENCY

As we know, using public key cause extra overloads on the system and reduces the performance of the system. Public key is widely used in anonymous systems and secure methods. In the proposed algorithm we use public key to encrypt data. To improve the performance and to reduce the overload of public key encrypt/decrypt process, we use the private key. One of the most important and critical issues in private key cryptography is that if a key is revealed by the intruder, he can compromise the messages which transferred. So protecting the key is too crucial. In the proposed algorithm, each peer has a table of private keys. Each entity of this table determines the private key that must be used when the peer wants to send a message to the other peer. If there are N peers in the network, the table has N-1 entities to store other peer's private keys. The biggest challenge in this algorithm is exchanging the private keys between the peers. The following method is used for exchanging the keys: The table is empty at first. When the peers want to send a message to the other peers, they look up the entities of those peers. If the entities are empty, they use public key to encrypt data and send their own private keys via encrypted data. When the peers receive a message, they extract private key of the senders and save it into the entities of those peers. If the entity of the destination peers is not empty, it will use this private key to encrypt the data. Although using these tables increase memory usage in each peer; however, reducing the overload of public key encryption process and response time has more benefit for these kinds of networks.

## VIII. THE ADVANTAGES AND THE DISADVANTAGES

The important issue in the proposed algorithm is suppernode. It causes single point of failure. To solve this problem we can use several suppernodes to cover each other if any suppernode downs. Each peer can connect to any suppernode and each suppernode synchronizes its list with others. Using several servers provide a network with more fault tolerance. As we mentioned before, the proposed algorithm uses several intermediate peers to connect requester to provider and change response/request path randomly. Therefore the proposed algorithm increases reliability, because if one intermediate peer suddenly leaves the network, the requester will choose another path to connect to provider. Also we can consider the network traffic to use the paths with less traffic in order to increase efficiency and reduce response time. One of the most important advantages of the proposed algorithm is its more prevention against traffic analysis, because the requester can change dual-path periodically so it is too difficult for the intruder to reveal the origin path of transferred data. Data sent via peer-to-peer communications is vulnerable to traffic analysis. Traffic analysis is the process of intercepting and analysing messages in order to compromise information from patterns in communications.

Also the intruder cannot use time-to-live attack against network, because each path has different time to live so the intruder cannot gather useful information to reveal requester.

## IX. CONCLUSION

The proposed Dual-Path peer-to-peer anonymous algorithm provides a flexible layer for the requester to choose dual-path to connect to the provider. This algorithm provides more reliability and more protection against traffic analysis and time-to-live attack. Also it increases its fault tolerance as the connection between the requester and the provider is not depend on intermediate peers, and if each intermediate peer downs, the requester can change the dual-path to continue its connectivity.

ACKNOWLEDGMENT

The authors would like to thank the "Education & Research Institute for ICT" for supporting us during doing this research.

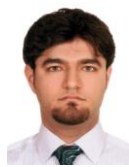
**Ehsan Saboori** is graduated from Ferdowsi University of Mashhad, Iran in computer engineering and is graduated from K.N Toosi University of technology, Tehran, Iran in IT engineering. He is interested in "Peer-to-Peer Networks", "Computer Networks", "Network Security" and "Anonymity". He currently works on peer to peer network security and privacy.